\documentclass[final,1p,times]{elsarticle} 
\usepackage{graphicx} 
\usepackage{amssymb} 
\usepackage{amsthm} 
\usepackage{lineno} 

\journal{Nuclear Physics A} 
\begin{document} 

\begin{frontmatter} 


\title{Non-photonic electron-hadron azimuthal correlation for AuAu,
CuCu and pp collisions at $\sqrt{s_{NN}}=200$ GeV}

\author{Bertrand Biritz$^{a}$ for the STAR collaboration}

\address[a]{University of California, Los Angeles, California 90095, USA}

\begin{abstract} 
We present preliminary STAR results of azimuthal correlations
between non-photonic electrons and hadrons in AuAu, CuCu and pp at $\sqrt{s_{NN}}=200$ GeV. Comparison of the e-h correlations from these colliding systems allows one to study the system-size dependence of heavy quark jet-medium interactions. We also report on the relative charm and bottom contributions to non-photonic electrons extracted from correlations measured in pp collisions. Our results, when combined with $R_{AA}$ measurements of non-photonic electrons, constrain the charm and bottom energy loss in the dense medium.
\end{abstract} 

\end{frontmatter} 



\section{Introduction}
Data for di-hadron correlations in AuAu and dAu using the STAR detector show
a suppression of high-$\textrm{p}_{t}$ hadron yields and modifications in the 
azimuthal correlation in central AuAu collisions \cite{PRL98, PRL96}. On the away-side of the trigger particles, the observed broadening in the correlations reflects the presence of strong jet-medium interactions, possibly a Mach cone effect \cite{ZYAM, Starter}. Non-photonic electron triggered particle correlations probe heavy quark jet-medium interactions. Electron-hadron (e-h) correlations on the near side, triggered by non-photonic electrons from charm and bottom decays, have different patterns due to the large mass difference between $D$ and $B$ mesons. Using e-h correlations from pp collisions in comparison with PYTHIA calculations, we can estimate the relative $D$ and $B$ contributions to the non-photonic electrons.
    
\section{Analysis Technique}
The data sets used were collected by STAR at RHIC in 2007 for AuAu, 2006 for CuCu and 2005 ($3<\textrm{p}_{t}<6$ GeV/c) and 2006 ($6<\textrm{p}_{t}<9$ GeV/c) for pp. The Time Projection Chamber (TPC) \cite{TPC}, the heart of STAR, was used to reconstruct the tracks of charged particles. Additionally the Barrel Electromagnetic Calorimeter (BEMC) and Shower Maximum Detector (SMD) \cite{Barrel}, the electromagnetic calorimeters which surround the TPC in full $2\pi$ azimuth, were required to have a least one track projected from the TPC onto it and also have an energy deposition greater than a predetermined minimum (i.e. high-tower threshold). This is to increase the number and purity of high $\textrm{p}_{\textrm{t}}$ electrons. For CuCu the threshold was at 3.75 GeV, 5.5 GeV for AuAu and 5.4 GeV for pp. For pp and AuAu the pseudorapidity used was $|\eta|<0.7$, while for CuCu it was $0< \eta<0.7$ due to the partial installation of the BEMC/SMD. For the AA collisions we used a 0-20\% centrality cut, the definition of which is explained in \cite{Centrality}.

To identify electron candidates one used a combination of measurements: ionization energy loss (dE/dx) in the TPC, ratio of particle momentum to energy deposited in the BEMC and lastly the electromagnetic shower size in the SMD---see \cite{Method, Dong} for more detail.

The primary background are photonic electrons either coming from photon conversions inside STAR or the Dalitz decays of $\pi^{0}$ and $\eta$. In both cases the electron pairs have a small invariant mass. This background is removed by pairing electron candidates with a track having passed a loose dE/dx cut around the electron ionization band. The distribution of 2D invariant masses ($\Delta\phi$ is ignored to minimize tracking resolution effects \cite{Dong}) for opposite sign ({\it OppSign}) pairs is obtained upon which a cut of $< 0.1$ GeV/$\textrm{c}^{2}$ is applied to reject most photonic pairs. The combinatorial background is estimated using the 2D invariant mass distribution of same sign ({\it SameSign}) pairs.

The azimuthal correlation of non-photonic electrons and hadrons begins with a semi-inclusive ({\it semi-inc.}) electron sample, which is the inclusive electron sample minus the {\it OppSign} background (after having applied the mass cut).

The correlation is done via the following method:
\begin{equation}
\Delta\Phi_{\textrm{non-photonic}}=\Delta\Phi_{\textrm{semi-inc}}-\Delta\Phi_{\textrm{not reco. photonic}}+\Delta\Phi_{\textrm{comb.}}
\end{equation}
\begin{equation}
\Delta\Phi_{\textrm{not reco. photonic}}=\Bigl(\frac{1}{\epsilon}-1\Bigr)\Delta\Phi_{\textrm{reco. photonic}}
\end{equation}
$\Delta\Phi_{\textrm{comb.}}$ is the same sign combinatorial background, while $\Delta\Phi_{\textrm{not reco. photonic}}$ are the photonic electrons which weren't reconstructed due to inefficiencies $\epsilon$, which is the photonic electron reconstruction efficiency estimated via simulations and found to be $\sim 60\%$ for AuAu, $\sim 66\%$ for CuCu and $\sim 70\%$ for pp. There is only a minor transverse momentum dependence of the efficiency in the momentum range being looking at. More details on this method using a semi-inclusive electron sample are in \cite{Electrons, Electrons2}.

\section{Results}
\begin{figure}[!h]
\centering
\includegraphics[scale=0.70]{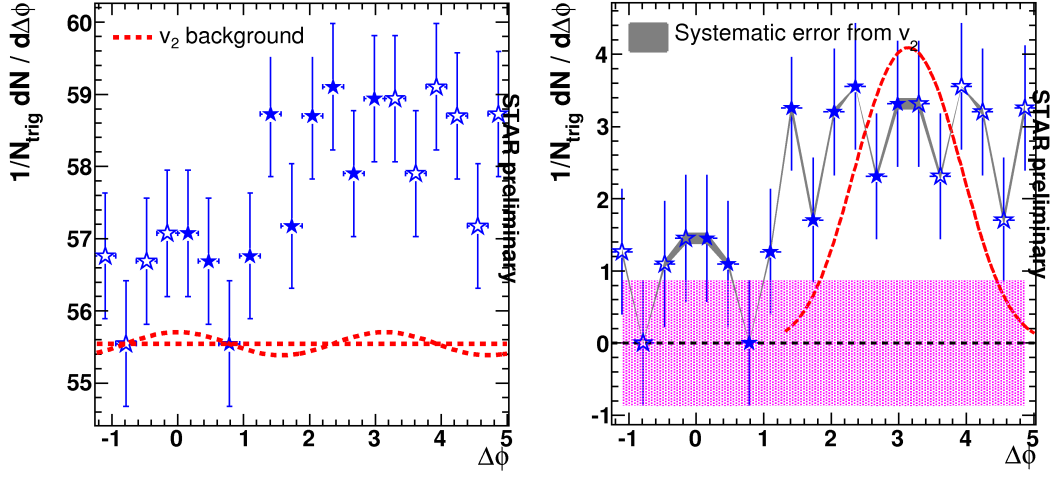}
\caption{Non-photonic e-h correlation in AuAu collisions at 200 GeV. Panel (a) shows the correlation before $\textrm{v}_{2}$ background subtraction and panel (b) after the subtraction, with a dashed fitting curve from PYTHIA pp expectations on the away side. The error bars are statistical, and the error band around zero shows the systematical uncertainty from ZYAM.\label{Gold}}
\end{figure}

\begin{figure}[!h]
\centering
\includegraphics[scale=0.70]{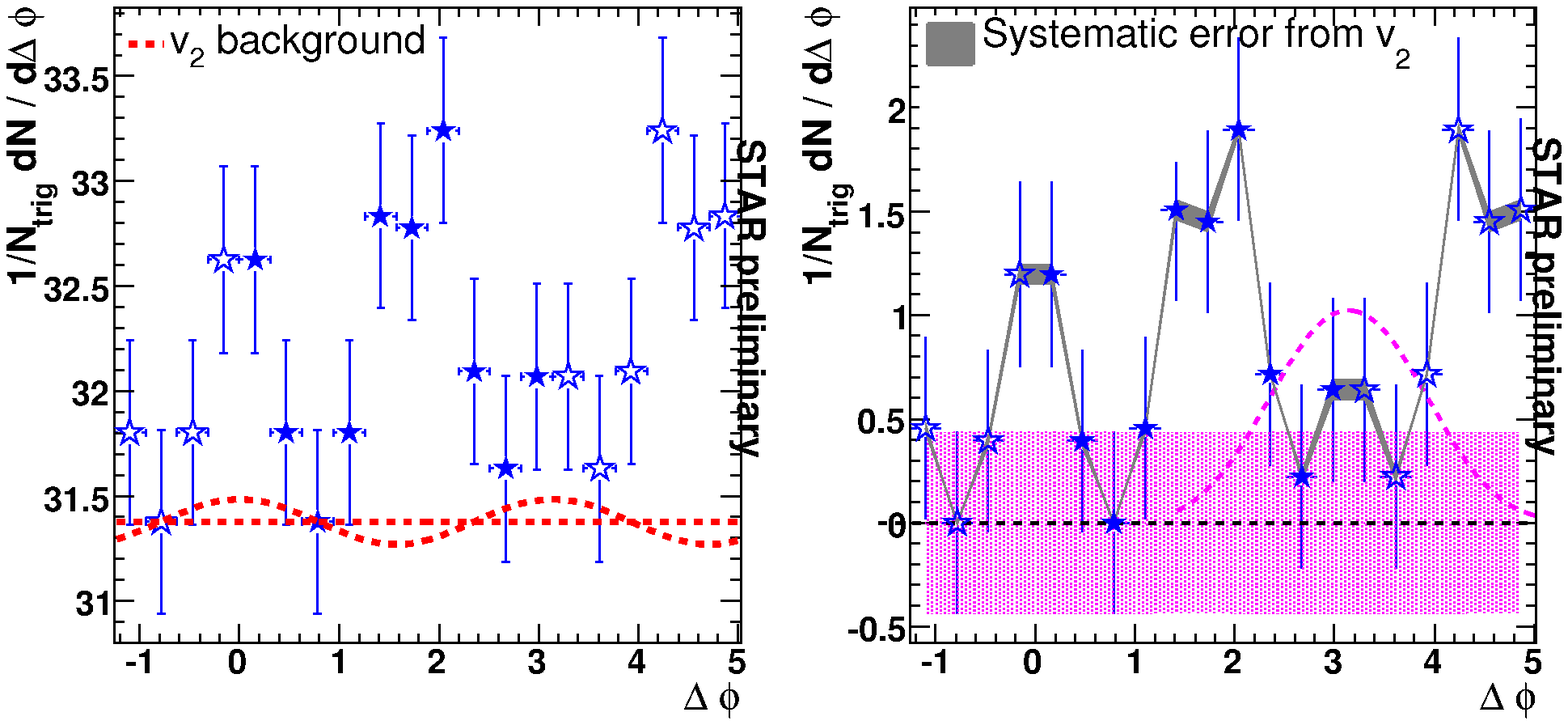}
\caption{Non-photonic e-h correlation in CuCu collisions at 200 GeV. Panel (a) shows the correlation before $\textrm{v}_{2}$ background subtraction and panel (b) after the subtraction, with a dashed fitting curve from PYTHIA pp expectations on the away side. The error bars are statistical, and the error band around zero shows the systematical uncertainty from ZYAM.\label{Copper}}
\end{figure}

Figures \ref{Gold} and \ref{Copper} plot the e-h azimuthal correlation for AuAu and CuCu at 200 GeV. The left panels show the raw correlations along with dashed curves for elliptic flow ($v_{2}$) \cite{v2} and  zero yield at minimum (ZYAM) \cite{ZYAM}. Systematic uncertainty of $v_{2}$ is determined by using a lower limit of zero and upper limit of 80\% for AuAu or 60\% for CuCu of charged hadron $v_{2}$. The momentum ranges used are 3 GeV/c $<\textrm{p}_{\textrm{t}}<6.0$ GeV/c for trigger electrons and 0.15 GeV/c $<\textrm{p}_{\textrm{t}}<0.5$ GeV/c for hadrons in AA collisions. To decrease the error bars, the correlation is folded into [0, $\pi$] and the data points beyond are reflections. One clearly already sees a modification of the away-side. The right panels plot the correlation after the $v_{2}$ subtraction and ZYAM application. On the away-side (around $\pi$) instead of a single peak there is a broadening (AuAu) or possible double-peak (CuCu) structure. Using a PYTHIA calculation for pp to fit this one sees the $\chi^{2}$/ndf is rather poor. This modification of the away-side is similar to the di-hadron case in AuAu \cite{Starter} and probably indicates heavy quark interaction with the dense medium.

\begin{figure}[!h]
\centering
\includegraphics[scale=0.35]{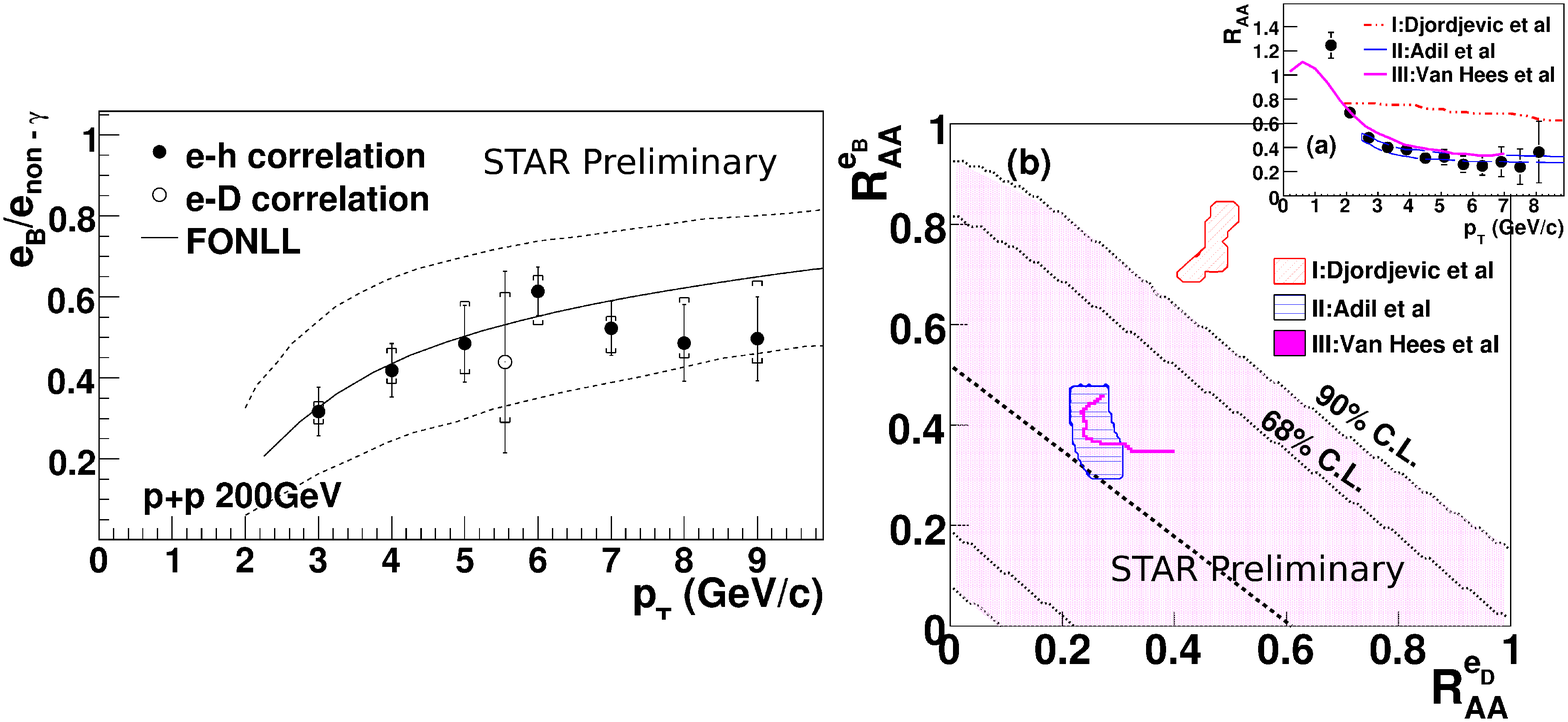}
\caption{(a) Transverse momentum dependence of the relative contribution from B mesons to the non-photonic electron yields. The solid curve illustrates the FONLL calculation \cite{FONLL}. Theoretical uncertainties are indicated by the dashed curves. e-D correlations are described in \cite{Andre1, Andre2} (b) Correlations of $R_{AA}$ for electrons from B meson ($R^{eB}_{AA}$) and D meson ($R^{eD}_{AA}$) decays for $p_{T}>5$ GeV/c. (c) $R_{AA}$ for the non-photonic electrons as a function of $p_{T}$ \cite{PRL98}.\label{Shingo}}
\end{figure}
Fitting the near-side of the e-h correlation over various trigger $\textrm{p}_{\textrm{t}}$ ranges with a PYTHIA simulation (fitting method is described in \cite{Electrons, Electrons2}) one sees equal charm and bottom contribution above 5 GeV/c, Figure \ref{Shingo} (a). Combining this with the previously measured $R_{AA}$ for non-photonic electrons (\cite{Method}, Figure \ref{Shingo} (c)) one has 

\begin{equation}
R^{\textrm{non-photonic}}_{AA}=(1-r_{B})R^{e_{D}}_{AA}+r_{B}R^{e_{B}}_{AA},\quad
r_{B}=\frac{e_{B}}{e_{B}+e_{D}}=\frac{e_{B}}{e_{non-\gamma}}
\end{equation}

Figure \ref{Shingo} (b) shows the most probable values for $R^{e_{D}}_{AA}$ and $R^{e_{B}}_{AA}$ with the dashed line being the 90\% confidence limit (taking $R_{AA}$ and $r_{B}$ uncertainties into consideration). Even if D decay electrons are fully suppressed $R^{e_{B}}_{AA}<1$. This indicates B meson yields are suppressed at high $\textrm{p}_{\textrm{t}}$ in heavy ion collisions presumably due to $b$ quark energy loss in the dense matter \cite{Model1}, modification of the fragmentation process due to the dense medium and/or dissociation/absorption of heavy flavor hadrons in the dense medium during the evolution \cite{Model2}. From it one can see that electrons coming from B decays are as suppressed as those from D decays. Models I, II and III are described in \cite{Model1}, \cite{Model2} and \cite{Model3} respectively.

\end{document}